\documentclass[12pt]{iopart}

\usepackage{graphicx}
\usepackage{xspace}
\usepackage{xcolor}

\definecolor{blendedred}{rgb}{0.8,0.2,0.2}
\definecolor{blendedgreen}{rgb}{0.2,0.8,0.2}

\newcommand{\eV}{\ensuremath{\mathrm{eV}}}

\begin{document}

\title[Robustness of topologically protected transport in 
graphene--hBN heterostructures]%
{Robustness of topologically protected transport in 
graphene--boron nitride lateral heterostructures}
\author{D~S~L~Abergel}
\address{Nordita, KTH Royal Institute of Technology and Stockholm
	University, Roslagstullsbacken 23, SE-106 91 Stockholm, Sweden.}
\address{Center for Quantum Materials, KTH and Nordita,
	Roslagstullsbacken 11, SE-106 91 Stockholm, Sweden.}

\begin{abstract}
	Previously, graphene nanoribbons set in lateral heterostructures
	with hexagonal boron nitride were predicted to support topologically
	protected states at low energy.
	We investigate how robust the transport properties of these states
	are against lattice disorder. We find that forms of disorder that do
	not couple the two valleys of the zigzag graphene nanoribbon do not
	impact the transport properties at low bias, indicating that these
	lateral heterostructures are very promising candidates for
	chip-scale conducting interconnects. 
	Forms of disorder that do couple the two valleys, such as vacancies
	in the graphene ribbon, or substantial inclusions of armchair edges
	at the graphene--hexagonal boron nitride interface will negatively
	affect the transport. 
	However, these forms of disorder are not commonly seen in current
	experiments.
\end{abstract}


\maketitle

\section*{Introduction}

If graphene devices are to be integrated into circuits with multiple
components, it will be highly convenient to have graphene connectors to
allow electronic current to move between different devices. 
Zigzag graphene nanoribbons (ZZGNRs) are known to host metallic edge
states, and are attractive candidates for such
current-carrying wires.
This idea has been investigated previously, but it was found that the
transport properties of the ZZGNRs were highly fragile against edge
roughness \cite{Gunlycke-APL90, Yoon-APL91, Basu-APL92, Cresti-NanoRes1, 
Martin-PRB79}. 
This is a crucial issue, since any growth-based fabrication method for
will necessarily introduce lattice scale disorder into the
ZZGNRs. 
Lateral heterostructures are monolayers where two or more 2D materials
are `stitched' together to form 1D interfaces. It is possible to grow
lateral heterostructures of graphene and insulating hexagonal boron
nitride (hBN) \cite{Levendorf-Nature488, Sutter-NanoLett12,
Liu-NatNano8, Liu-Science343, Han-ACSNano7, Gao-NanoLett13,
Drost-NanoLett14}, and much theoretical work, especially using
\textit{ab initio} methods, has been done to investigate their
electronic properties
\cite{Ding-APL95, Bhowmick-JPCC115, Jungthawan-PRB84, Seol-APL98,
Li-PRB88}.
However, the topological properties of these lateral heterostructures
has received only very limited attention \cite{Jung-NanoLett12}.
Jung \textit{et al.}~suggested that by engineering hBN ``cladding'' on
either side of the ZZGNR, then the topological properties of the
combined system can ensure that there are always
conducting channels in the graphene \cite{Jung-NanoLett12}.
This topological protection is akin to the Jackiw-Rebbi states that are
predicted to exist at mass-inversion boundaries in hexagonal crystals
\cite{Abergel-NJP16}, and is described by a valley Chern number
\cite{Jung-NanoLett12,Li-PRB82}.

The atomic-scale precision required to produce these lateral
heterostructures does currently exist \cite{Gao-NanoLett13,
Drost-NanoLett14}, but the graphene-hBN interfaces defined by such
techniques are still rather disordered.
This issue is crucial in the context of topological protection because,
as explained in Refs.~\cite{Jung-NanoLett12, Li-PRB82} the topological
protection described by the valley Chern number only persists as long as
the two valleys in the graphene are not coupled to each other. 
However, the edge roughness may constitute short-range disorder that is
strong enough to scatter electrons between the two valleys, breaking the
topological protection and allowing backscattering which reduces the
conductance.

In this manuscript, we present a full analysis of the role of
atomic-scale disorder on the transport properties of ZZGNR-hBN lateral
heterostructures.
We find that only certain types of lattice disorder break the
topological protection and couple the valleys, allowing the
backscattering.
Specifically, inclusion of substantial regions of armchair interface and
graphene vacancies will do this, but rough edges and inclusions of
random boron or nitrogen atoms in the graphene ribbon will not. 
We contend that the explanation for this is that the latter types of
disorder are smooth enough in the sense that they can be reached from
the original Hamiltonian by an adiabatic transformation, and therefore
they do not modify the overall topological properties of the system and
hence do not couple the valleys.
This analysis, combined with recent advances in fabrication techniques,
reopens the issue of ZZGNRs clad with hBN as a highly suitable method of
providing chip-scale conducting channels.

\section*{Methods}

\begin{figure}[t]
	\centering
	\includegraphics[]{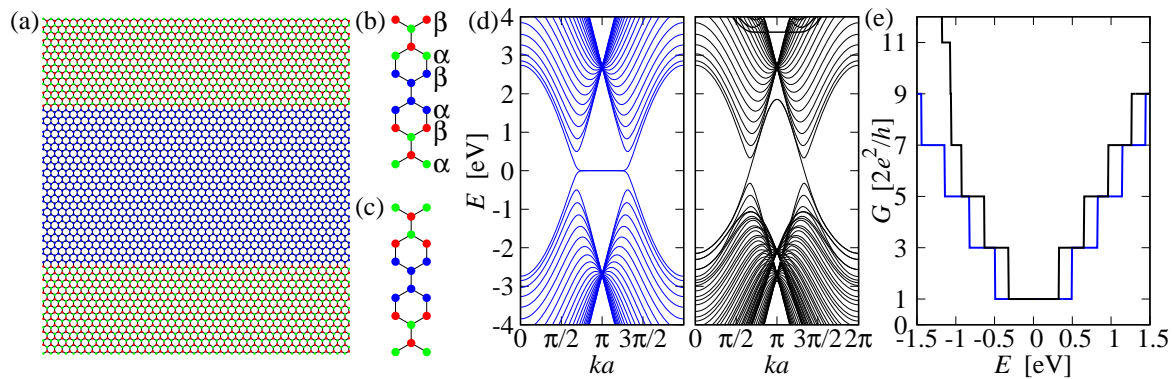}
	\caption{(a) The non-disordered ZZGNR with hBN cladding.
	Atomic sites are colour-coded as follows.  Carbon: blue; nitrogen:
	red; boron: green.
	(b) Sketch of the `same' topology configuration. Notice that the
	$\alpha$ site both above and below the ZZGNR hosts a boron atom.
	(c) Sketch of the `opposite' topology configuration. In this case,
	the $\alpha$ site below the ZZGNR hosts a boron atom, while above the
	ZZGNR, the $\alpha$ site hosts a nitrogen atom.
	(d) Band structures of non-disordered $5\mathrm{nm}$ ZZGNR without
	(blue) and with (black) hBN cladding.
	(e) Conductance of non-disordered $5\mathrm{nm}$ ZZGNR without
	(blue) and with (black) hBN cladding.
	\label{fig:cleanwire}}
\end{figure}

To demonstrate the robustness of the predicted low bias transport
channel, we compute the 1D charge conductance using the
Landauer-Buttiker scattering formalism within the Kwant package
\cite{Groth-NJP16}. 
We use a ZZGNR with width $5.0\mathrm{nm}$ and length $10.0\mathrm{nm}$,
corresponding to chip-scale dimensions, and we take the hBN width to be
$3.0\mathrm{nm}$, which is wide enough to ensure the topological
properties are manifested whilst still being small enough to ensure
reasonable computation time. This configuration is shown in
Fig.~\ref{fig:cleanwire}(a).
To compute the spectrum and wave functions in the leads and in the
scattering region, we use a nearest-neighbour tight binding model. 
The tight binding parameterisation requires the onsite energies $U_i$
for each chemical species, and the hopping elements $t_{ij}$ between 
them, where $i,j\in \{\mathrm{C}, \mathrm{B}, \mathrm{N}\}$ denote the
chemical species.
Throughout, we use $U_{\mathrm{C}} = 0$,
$U_{\mathrm{B}}=3.6\eV$, $U_{\mathrm{N}} = -1.0\eV$,
$t_{\mathrm{CC}} = 2.7\eV$, $t_{\mathrm{CB}} = 2.1\eV$, $t_{\mathrm{CN}}
= 2.3\eV$, and $t_{\mathrm{BN}} = 2.5\eV$ \cite{Jungthawan-PRB84}. 
We stress that so long as $\mathrm{signum}(U_\mathrm{N}) \neq
\mathrm{signum}(U_\mathrm{B})$, the precise values of the tight binding
parameters do not make any difference to the overall topological
properties of the system, and merely give small quantitative changes to
the band structure and hence the exact positions of the conductance
steps.

In principle, a more accurate description of the band structure of the
ribbons is given by a third-nearest neighbour tight binding theory
\cite{Hancock-PRB81, Wu-NanoResLett6}, but this additional complexity
changes none of the qualitative features of the results, or the
considerations about the topology of the system. 
Therefore, we restrict our discussion to the nearest neighbour model for
clarity.
In the Supplementary Material, we show data which justifies this
assumption further.

\section*{Results and discussion}

Figure \ref{fig:cleanwire}(a) shows the non-disordered ZZGNR with hBN
cladding that forms the basis of the ribbons we consider in this
manuscript. 
In the `same' topology configuration, sketched in
Fig.~\ref{fig:cleanwire}(b), the $\alpha$ and $\beta$ sublattices of the
two regions of hBN cladding have the same chemical orientation.
(Throughout, we use the notation $\alpha$ and $\beta$ for the two
sublattices of the hexagonal crystal to avoid confusion between the `B'
sublattice and the chemical symbol for boron atoms.) 
In this case, there are no topologically protected states in the
graphene because the mass gap generated in the hBN has the same sign in
both hBN regions. 
However, when the chemical orientation is `opposite', sketched in
Fig.~\ref{fig:cleanwire}(a) and Fig.~\ref{fig:cleanwire}(c), the mass
term has opposite sign in the two hBN regions, and therefore their
valley Chern number is different. 
This is sufficient to ensure that there are topologically protected
states in the graphene \cite{Jung-NanoLett12}.

\begin{figure}[t]
	\centering
	\includegraphics[]{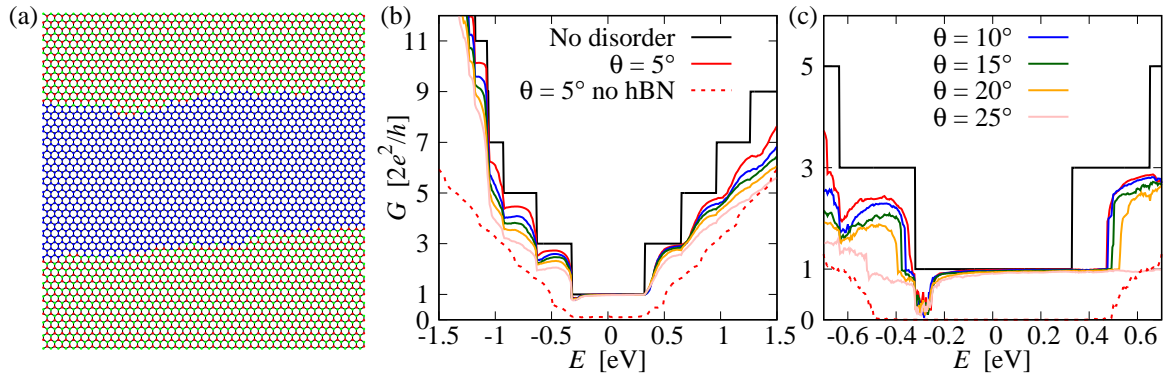}
	\caption{(a) Typical disordered wire with $\theta=15^{\circ}$.
	(b) Averaged conductance, and (c) fifth percentile conductance
	of 200 disordered wire realisations.
	\label{fig:randedge}}
\end{figure}

The band structures and conductance of the ZZGNR are shown in
Figs.~\ref{fig:cleanwire}(d),(e), respectively, where the blue lines
correspond to the ZZGNR with no cladding and the black lines are for
cladding in the `opposite' topology configuration with nitrogen
termination. These are fully consistent with the results of
Ref.~\cite{Jung-NanoLett12}.
The units on the vertical axis are $2e^2/h$ to account for the spin
degeneracy of each of the bands.
We see that the cladding causes the flat band edge states to be removed
and instead there are dispersing modes at zero energy.
These are the topologically protected Jackiw-Rebbi-like modes, and they
manifest in the conductance by a finite minimum conductance plateau
$-0.35\mathrm{eV} < E < 0.35\mathrm{eV}$.
This is the feature that we are most interested in, since it defines the
conducting channel that may be used to direct current between graphene
devices.
The sharp increase in the conductance at roughly $E=-1\mathrm{eV}$ is
caused by conduction through modes located mainly on the nitrogen atoms
which can be seen in the band structure in Fig.~\ref{fig:cleanwire}(d).
Modes located mainly on the boron atoms are located at approximately
$E=3.6\mathrm{eV}$ and so do not appear in the conductance plot. The
presence of the cladding also slightly reduces the gap between the first
non-topologically protected modes in the conduction and valence bands.

Our central result is shown in Fig.~\ref{fig:randedge}, where we
plot the conductance of ZZGNRs with random edge disorder and hBN
cladding.
To create disordered ribbons, we keep track of the two boundaries
between the ZZGNR and the hBN. For each lattice unit cell
along the length of the wire, we allow the $y$ coordinate of the
boundaries to change relative to the previous unit cell $y_i =
y_{i-1}+\delta y$ where $\delta y = a \tan\theta_i$ and $a$ is the lattice
constant. The deflection angle $\theta_i$ is a random variable
characterised by a Gaussian distribution with variance $\theta$, so
that higher values of $\theta$ correspond to a higher propensity towards
rough edges. 
Figure \ref{fig:randedge}(a) shows a typical example ribbon with
$\theta=15^\circ$, and we stress that the shape of the upper and lower
boundaries are independent of each other.
For each value of $\theta$ we compute the conductance of 200 ribbons
with random disordered edges, calculate the mean conductance at each
value of energy, and show the result in Fig.~\ref{fig:randedge}(b). 
For even high values of edge roughness, on average, the finite minimum
conductance at low energy remains intact, indicating that the
topologically protected modes in the graphene are resiliant against edge
disorder. 
The reduction in the conductance at higher energy (\textit{i.e.}~away
from the finite minimum conductance plateau) is caused by backscattering
in the non-topologically protected states as would be expected in ZZGNRs
without hBN cladding \cite{Basu-APL92}.
To confirm the effect of the cladding, the red dotted line shows
the averaged conductance for the exact same 200 ZZGNRs with
$\theta=5^\circ$ but with the hBN cladding removed, and therefore with
no topological effects. 
In this case, the finite minimum conductance plateau at low energy is
completely absent.
In Fig.~\ref{fig:randedge}(c) we show the fifth percentile conductance 
at each energy, \textit{i.e.} the level at which 190 of the 200 disorder
realisations have better conductance than the line shown.
This shows that for the vast majority of randomly disordered ZZGNRs, the
finite minimum conductance plateau remains intact.

\begin{figure}[t]
	\centering
	\includegraphics[]{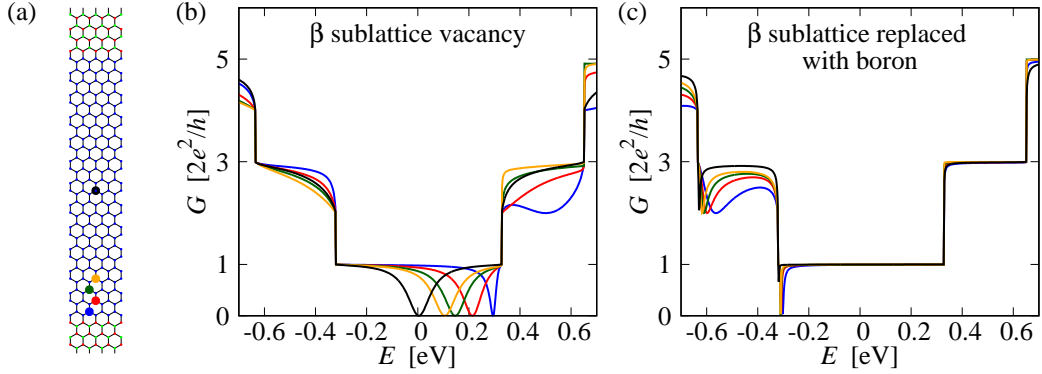}
	\caption{Single site disorder. (a) Sketch of hBN-clad ZZGNR showing
	the position of the vacancy or substitution site by dots
	corresponding in color to the lines in (b) and (c).
	(b) Conductance with graphene vacancy on the $\beta$ sublattice. 
	(c) Conductance with boron substitution on the $\beta$ sublattice.
	\label{fig:sitedisorder}}
\end{figure}

We now examine some more controlled forms of disorder to determine which 
contribute to the breakdown of the topological protection. 
Figure \ref{fig:sitedisorder} shows the calculated conductance for
carbon vacancies and boron substitution in the ZZGNR.
In principle, since this type of disorder has the shortest associated
length scale, it should scatter electrons between states separated by a
momentum of the order of the Brillouin zone size, and therefore couple
the $K$ and $K'$ valleys the most strongly, leading to breakdown of the
topological protection of the low energy modes.
In Fig.~\ref{fig:sitedisorder}(b), we show the conductance of our
standard ZZGNR with a single carbon atom removed from the $\beta$ 
sublattice of a unit cell near the lower edge. The location of the
vacancies are shown in panel (a), where the colour of the dot
corresponds to the colour of the line in the conductance plot.
If it is removed from the first unit cell (blue line), then a sharp
decrease in the conductance at the top of the low
energy finite conductance plateau is seen. As the vacancy is moved into
the ZZGNR (red, green, and orange lines) the position of the
resonance moves towards zero energy. 
When the vacancy is in the center of the ZZGNR (black line), the
conductance dip is at the center of the plateau. 
This variation in the energy at which the conductance dip occurs with
the position of the vacancy is related to the change in the wave
function with energy, since modes with energy near the edge of the
finite minimum conductance plateau are localised near the edges of the
ZZGNR.
The mirror symmetry of the opposite topology configuration insists that
vacancies on the $\alpha$ sublattice at the upper edge have their
conductance dip at the same energy. Swapping either the sublattice or
the edge makes the sign of the energy at which the conductance dip
occurs change.

In contrast, Fig.~\ref{fig:sitedisorder}(c) shows that substitution of a
carbon atom with a boron atom has very little effect on the conductance.
We have verified that nitrogen substitution gives even smaller changes
to the conductance.
This explains the small dip in the conductance in
Fig.~\ref{fig:randedge} at the low energy side of the finite minimum
conductance plateau, since mild edge roughness is replicated by many
boron or nitrogen substitutions.

\begin{figure}[t]
	\centering
	\includegraphics[]{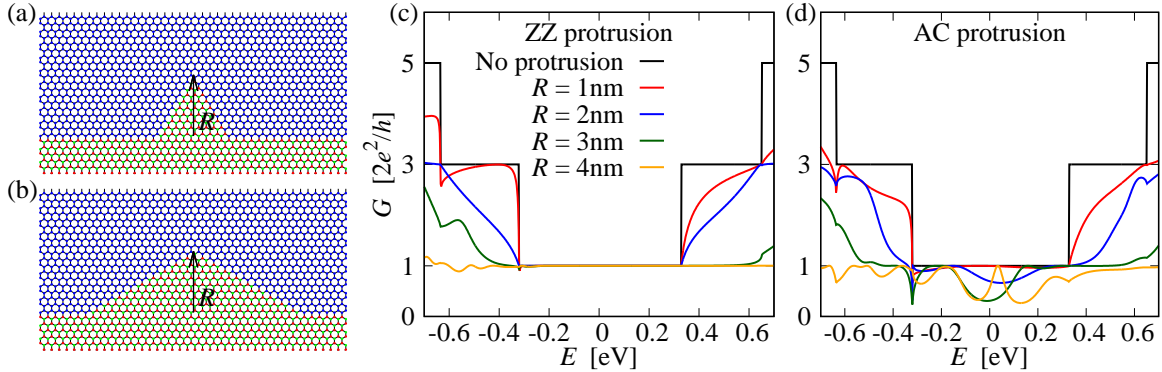}
	\caption{Protrusions with ordered edges. (a) Sketch of a zigzag
	protrusion. (b) Sketch of an armchair protrusion. (c)
	Conductance of hBN-clad ZZGNR with zigzag protrusion.
	(d) Conductance of hBN-clad ZZGNR with armchair protrusion.
	\label{fig:protrusions}}
\end{figure}

In Fig.~\ref{fig:protrusions} we demonstrate the crucial difference
between zigzag and armchair protrusions of hBN into the ZZGNR. It is
well known that armchair edges strongly couple the two valleys, since
the projection of the 2D Brillouin zone of bulk graphene onto the 1D
Brillouin zone of a ribbon in the armchair direction projects the $K$
and $K'$ valleys to the same point. 
In contrast, the zigzag ribbon projects the $K$ and $K'$ valleys to
different points in the 1D Brillouin zone and so the two valleys are not
strongly coupled (see the Supplementary Material).
Figures \ref{fig:protrusions}(a) and (b) show a triangular protrusion of
hBN into the ZZGNR with width $R$ with zigzag and armchair edges,
respectively. Figure \ref{fig:protrusions}(c) shows the calculated
conductance for the zigzag protrusion up to $R=4\mathrm{nm}$ (i.e.
$80\%$ of the ZZGNR width). The finite conductance minimum associated
with the topological states is perfectly intact for all zigzag
protrusions, indicating that there is no backscattering induced
in this case. 
However, backscattering is induced in the high energy non-topologically
protected states by the zigzag protrusion.
Figure \ref{fig:protrusions}(d) shows the
equivalent data for armchair protrusions. In this case, the valley
coupling manifests as substantial and apparently uncontrolled
oscillations in the conductance as a function of energy in the finite
minimum conductance plateau.
This indicates that substantial inclusions of armchair edges in the
ZZGNR lead to sub-optimal low energy transport properties.

\section*{Conclusions}

We have shown highly promising results for the use of hBN-clad ZZGNRs
as chip-scale interconnects with perfect ballistic conductance.
In particular, we have systematically investigated the impact of lattice
disorder of various types on the transport properties of such ZZGNRs.
Our conclusion is that only atomic-scale disorder which strongly couples
the $K$ and $K'$ valley of the bulk graphene --- thus invalidating the
valley Chern number construction and hence lifting the topological
protection of the low energy modes --- is effective at inducing
backscattering between the topological states and weakening the
transport.
Other types of disorder, such as edge roughness and chemical
substitution, do not impact the transport. 
This is because, at least within a tight binding model, atomic
substitution amounts only to changing the onsite energy and hopping
elements by a finite amount.
This is a only a small change in the theory from the `clean' case, or,
in the language of topology, it is an adiabatic transformation and hence
retains the same topological properties.
This is a very encouraging result, since zigzag interfaces are
energetically favourable to armchair interfaces, and have been shown to
dominate in fabrication by a ratio of better than 3:1, and carbon
vacancies are very rare \cite{Gao-NanoLett13}. 
However, devices do exhibit substantial edge roughness. 
Therefore, topologically protected transport may be expected in such
devices even with current growth techniques.

One issue that has so far not been addressed is that of the $1.8\%$
lattice mismatch between graphene and hBN. 
However, the results presented here are enough to justify that this
should not present a problem for short interconnects. 
For ZZGNRs which are less than approximately 50 unit cells in length
(NB, the one we have modelled here is 40 unit cells long), the strain
buildup in the hBN due
to the mismatch should result only in the modification of the hopping
parameters between the last few rows of BN lattice sites and the first
few rows of carbon lattice sites. However, our results indicate that
small modifications in the
hopping parameters do not result in changes in the topological
properties of the system and hence there should be no impact on the
topologically protected transport channels. For longer wires, we have
explicitely calculated the conductance with a missing nitrogen atom at
the edge (see Supplementary Material), and found that there was no
impact on the topological modes in the graphene because there is
essentially no wave function weight in the hBN for energies
$U_\mathrm{N} < E < U_\mathrm{B}$. Hence,
the strain induced by the lattice mismatch for chip-scale ZZGNRs should
have minimal impact on the transport properties.


\section*{Acknowlegements}

This work is supported by Nordita and by ERC project DM-321031.

\section*{References}

\bibliographystyle{./jpcm.bst}
\bibliography{./MainDatabase,./GrapheneBNHetero,./mypapers}

\newpage

\setcounter{figure}{0}

\title[Supplementary material]%
	{Supplementary material to `Robustness of topologically protected
transport in graphene--hBN heterostructures'}%
\author{D~S~L~Abergel}
\address{Nordita, KTH Royal Institute of Technology and Stockholm
	University, Roslagstullsbacken 23, SE-106 91 Stockholm, Sweden.}
\address{Center for Quantum Materials, KTH and Nordita,
	Roslagstullsbacken 11, SE-106 91 Stockholm, Sweden.}

\maketitle

\section*{Third nearest neighbour tight binding model}

In principle a third-nearest neighbour (3NN) tight binding model is more
accurate than the simple nearest neighbour model we use in the main text
\cite{Hancock-PRB81, Wu-NanoResLett6}. 
Below, we provide justification for this assumption by comparing tight
binding calculations for ZZGNR-hBN lateral heterostructures.

To construct a 3NN model, successive neighbour hops are given by the
formula
\begin{equation*}
	t_{ij}(R) = A_{ij} e^{-R/\xi}
\end{equation*}
where $i,j\in\{\mathrm{C},\mathrm{B},\mathrm{N}\}$ denote the the atomic
species, $\xi=0.4329$ is the decay constant, and we calculate $A_{ij}$
from the nearest-neighbour hops which are taken from
Ref.~\cite{Jungthawan-PRB84}. We emphasize that the precise values of
the hopping parameters does not have a qualitative effect on the band
structures or topology.

\begin{figure}[t]
	\centering
	\includegraphics[]{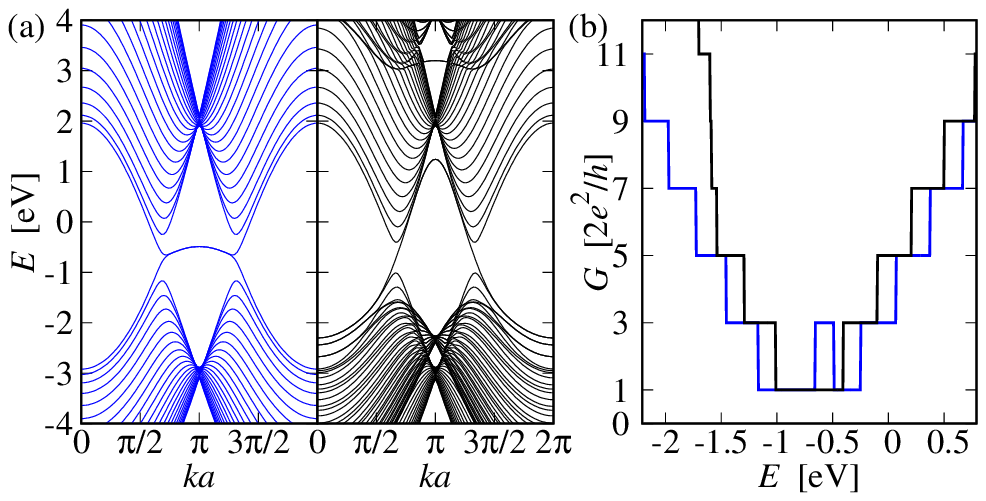}
	\caption{Equivalent data to Figure 1 of the main text for a
	third-nearest neighbour tight binding model.
	\label{fig:3nncleanwire}}
\end{figure}

In Fig.~\ref{fig:3nncleanwire}, we show the band structure and conductance
of a clean ZZGNR with hBN cladding in the 3NN tight binding
approximation. 
The inclusion of the second-nearest neighbour terms are especially
important since they introduce an effective onsite element. 
This includes a constant term which gives a rigid shift of the bands
relative to the nominal zero of the energy axis. This is not physically
meaningful. It also contains a contribution proportional to $k^2$
which introduces curvature to the previously flat band edge states when
the hBN cladding is absent. 
This leads to an additional step to $G=6e^2/h$ at $E \approx
0.5\mathrm{eV}$ in the finite minimum conductance plateau.

When the hBN cladding is present, the edge states disappear and are
replaced by the dispersing topologically protected states, just as in
the nearest-neighbour model. The conductance plot for the 3NN case,
shown in Fig.~\ref{fig:3nncleanwire}(b), demonstrates that the conductance
with the hBN cladding is almost identical to that of the
nearest-neighbour case once the energy shift is accounted for.
This justifies our assumption to limit the main text to the nearest
neighbour model for clarity.

\begin{figure}[t]
	\centering
	\includegraphics[]{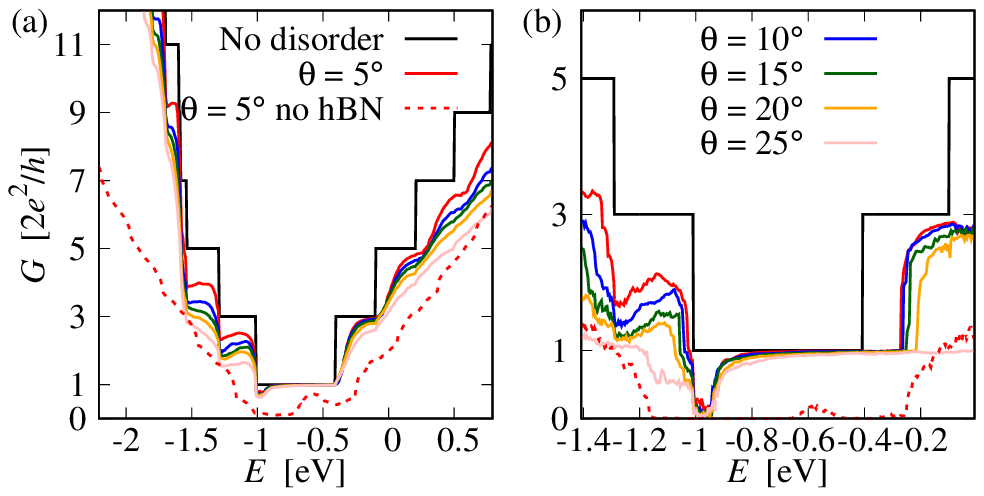}
	\caption{Equivalent data to Figure 2 of the main text for a
	third-nearest neighbour tight binding model.
	\label{fig:3nnrandedge}}
\end{figure}

In Fig.~\ref{fig:3nnrandedge} we plot the conductance data for the 3NN
model that corresponds to Fig.~2 of the main text. We stress that the
exact same 200 disorder realisations were used in both cases and the
only difference between the calculations is the number of hopping terms
in the tight binding theory. 
The qualitative features of the conductance in the 3NN case are
identical to the nearest-neighbour approximation and all conclusions
remain unchanged.

\section*{Nitrogen vacancy at ZZGNR-hBN interface}

\begin{figure}[t]
	\centering
	\includegraphics[]{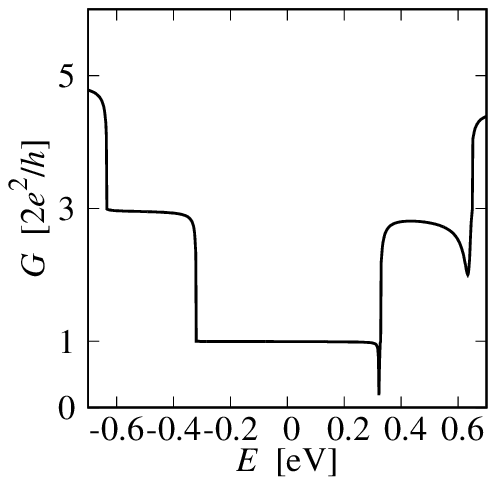}
	\caption{Conductance with nitrogen atom removed from ZZGNR-hBN
	interface to simulate the effect of strain relaxation in longer
	ribbons.
	\label{fig:nitremsite}}
\end{figure}

Figure \ref{fig:nitremsite} shows the conductance of a ZZGNR-hBN lateral
heterostructure where one of the nitrogen sites in the terminating row
of hBN has been removed. In conjuction with the observation that
`adiabatic' changes in the parameters of the hopping theory do not
introduce backscattering, the almost perfect finite minimum conductance
plateau in the figure shows that relaxation of strain buildup by a
missing nitrogen line does not negatively impact the transport in the
topologically protected ZZGNR states. This is because, at the energy of
relevance, the wave function of the topologically protected states is
located in the graphene and not on the boron or nitrogen atoms.

\section*{One-dimensional Brillouin zones}

\begin{figure}[tb]
	\centering
	\includegraphics[]{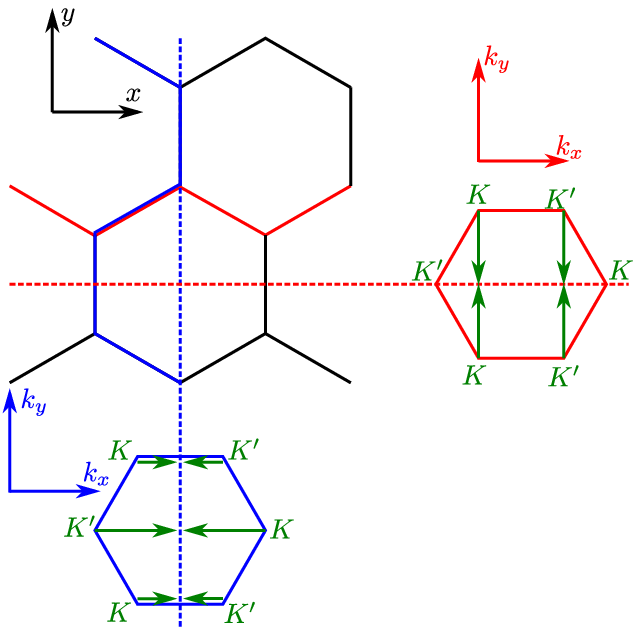}
	\caption{Projections of the 2D Brillouin zone into 1D for zigzag
	ribbons (red) and armchair ribbons (blue).
	\label{fig:BZProjection}}
\end{figure}

Figure \ref{fig:BZProjection} illustrates why armchair edges couple the
two valleys in graphene nanoribbons, while zigzag edges do not.
The hexagonal graphene lattice (shown in black, upper left) has a
2D Brillouin zone (BZ). 
When a ribbon is formed, the 1D translation symmetry
manifests as a 1D BZ, which can be formed from the 2D BZ by
projecting all momentum states down to the 1D BZ in the appropriate way.

For a zigzag edge (shown by the red line), the $x$ direction retains its
translational symmetry, and so $k_x$ is still a good quantum number.
Therefore, all $k_y$ states are projected down to $k_y=0$, as shown by
the green arrows on the red BZ sketched in the upper-right part of the
figure. This projection brings the $K$
and $K'$ points to different parts of the 1D BZ, and hence the valleys
are not coupled.

In contrast, for an armchair edge (shown by the blue line), the $y$
direction retains its translational symmetry and so $k_y$ remains a good
quantum number. 
The $k_x$ states are therefore brought to $k_x=0$, and as shown in the
lower-left part of the figure, the $K$ and $K'$ points arrive at the
same point in the 1D BZ. This couples the valleys strongly, and
invalidates the valley Chern number construction \cite{Li-PRB82}.

\end{document}